\documentclass[12pt]{article}

\usepackage{amsthm}
\usepackage[sumlimits]{amsmath}
\usepackage{amssymb}
\usepackage{algorithm}
\usepackage{algorithmic}
\usepackage[all]{xy}
\usepackage{wasysym}
\usepackage{fullpage}
\usepackage{graphicx}
\usepackage{url}
\usepackage{natbib}
\usepackage{authblk}
\usepackage{color}

\makeatletter
\newtheorem*{rep@theorem}{\rep@title}
\newcommand{\newreptheorem}[2]{%
\newenvironment{rep#1}[1]{%
 \def\rep@title{#2 \ref{##1}}%
 \begin{rep@theorem}}%
 {\end{rep@theorem}}}
\makeatother

\newtheorem{defn}{Definition}

\newreptheorem{lemma}{Lemma}
\newtheorem{thm}{Theorem}
\newreptheorem{theorem}{Theorem}

\theoremstyle{definition}

\newcommand{\cX}{\mathcal{X}}

\newcommand{\cA}{\mathcal{A}}

\begin{document}

\def\name{}
\def\email{\small\hfill\sc}
\def\addr#1{\small\it #1}

\title{
\bf\LARGE\hrule height5pt \vskip 4mm
Sparse Sequential Dirichlet Coding
\vskip 4mm \hrule height2pt
}

\renewcommand\Authsep{~~~~}
\renewcommand\Authand{~~~~}
\renewcommand\Authands{~~~~}
\setlength{\affilsep}{0.5em}

\author[$\dagger$]{Joel Veness}
\author[$\ddagger$]{Marcus Hutter}
\affil[$\dagger$]{\small University of Alberta, Edmonton, Canada}
\affil[$\ddagger$]{\small Australian National University, Canberra, Australia}

\maketitle

\vspace{-2em}
\begin{abstract}
This short paper describes a simple coding technique, Sparse Sequential Dirichlet Coding, for multi-alphabet memoryless sources.
It is appropriate in situations where only a small, unknown subset of the possible alphabet symbols can be expected to occur in any particular data sequence.
We provide a competitive analysis which shows that the performance of Sparse Sequential Dirichlet Coding will be close to that of a Sequential Dirichlet Coder that knows in advance the exact subset of occurring alphabet symbols.
Empirically we show that our technique can perform similarly to the more computationally demanding Sequential Sub-Alphabet Estimator, while using less computational resources.
\end{abstract}

\section{Introduction}

Suppose we needed to code a sequence of symbols $x_{1:n} := x_1 x_2 \dots x_n$ from an unknown alphabet $\cA$ generated by an unknown memoryless data generating source $\mu$.
If we knew an alphabet $\cX$ such that $\cA \subseteq \cX$, one solution would be to code the sequence using the Sequential Dirichlet Estimator 
\begin{equation}
\label{eq:sequential_dirichlet_def}
\rho_{\cX}(x_{1:n}) := \prod\limits_{i=1}^n \frac{c(x_{1:i}) + \tfrac{1}{2}}{i + \tfrac{|\cX|}{2} - 1},
\end{equation}
where $c(x_{1:n}) := \sum_{i=1}^{n-1} \mathbb{I}[x_n = x_i]$, as suggested by \cite{krichevsky1981pue}.
This technique has the property \citep{Tjalkens93sequentialweighting} that
\begin{equation}
\label{eq:dirichlet_bound}
-\log_2 \frac{\rho_{\cX}(x_{1:n})}{\mu(x_{1:n})} \leq \frac{|\cX| - 1}{2} \log_2 n + |\cX| -1.
\end{equation}
As Equation \ref{eq:dirichlet_bound} suggests however, performance of this particular coding technique can be poor for small values of $n$ when $|\cA|$ is much less than $|\cX|$.
This problem occurs often when using context-based techniques for data compression.
This is because, for many contexts, only a small subset of the full alphabet symbols are possible.
For example, when modeling English text it is very rare to see any character other than the letter \emph{u} immediately following the letter \emph{q}.
If we knew $\cA$ in advance, we could code $x_{1:n}$ using $\rho_{\cA}$, which from Equation \ref{eq:dirichlet_bound} would of course give a redundancy no greater than
\begin{equation}
\label{eq:small_alphabet_dirichlet_bound}
\frac{|\cA| - 1}{2} \log_2 n + |\cA| -1.
\end{equation}

The Sequential Sub-alphabet estimator proposed by \cite{Tjalkens93sequentialweighting} provides a natural Bayesian solution to this dilemma.
Rather than using the superset alphabet $\cX$, their technique weights over the set of all possible Sequential Dirichlet Estimators whose alphabets are \emph{subsets} of $\cX$.
This leads to an elegant algorithm that has a coding redundancy no more than
\begin{equation}\label{eq:redun_seqsubalpha}
\log_2 |\cX| + \log_2 {|\cX| \choose |\cA|} +  \frac{|\cA| - 1}{2} \log_2 n + |\cA| +1,
\end{equation}
when using a uniform prior over sub-alphabets.
Unfortunately this method requires $O(|\cX|)$ time to process each new symbol, and $O(|\cX|)$ space.
This can be prohibitive in situations where $|\cX|$ is large.
It would be better if the the time and space complexity were instead dependent on at most $|\cA|$.
This paper introduces a simple method, the Sparse Sequential Dirichlet Estimator, which achieves similar redundancy properties to the Sequential Sub-alphabet Estimator whilst being able to process each symbol in $O(1)$ time using at most $O(|\cA|)$ space.
 
\section{Preliminaries}

We begin with some notation for data generating sources.
An alphabet is a finite, non-empty set of symbols, which will denote as either $\cA$ or $\cX$. 
A binary string $x_1x_2 \ldots x_n \in \cX^n$ of length $n$ is denoted by $x_{1:n}$.
The prefix $x_{1:j}$ of $x_{1:n}$, $j\leq n$, is denoted by $x_{\leq j}$ or $x_{< j+1}$.
The empty string is denoted by $\epsilon$.
The concatenation of two strings $s$ and $r$ is denoted by $sr$.

A probabilistic data generating source $\rho$ is defined to be a sequence of probability mass functions $\rho_n : \cX^n \to [0,1]$, for $n\in\mathbb{N}$, satisfying the constraint that 
$$\rho_n(x_{1:n}) = \sum_{y\in\cX} \rho_{n+1}(x_{1:n}y)$$
for all $x_{1:n} \in \cX^n$, with base case $\rho_0(\epsilon) = 1$.
As the meaning is always clear from the argument to $\rho$, we drop the subscripts on $\rho$ from here onwards.
Under this definition, the conditional probability of a symbol $x_n$ given previous data $x_{<n}$ is defined as $\rho(x_n | x_{<n}) := \rho(x_{1:n}) / \rho(x_{<n})$ if $\rho(x_{<n}) > 0$, with the familiar chain rule $\rho(x_{1:n}) = \prod_{i=1}^n \rho(x_i | x_{<i})$ now following.

A source code $c : \cX^* \to \cX^*$ assigns to each possible data sequence $x_{1:n}$ a binary codeword $c(x_{1:n})$ of length $\ell_c(x_{1:n})$.
The typical goal when constructing a source code is to minimize the lengths of each codeword while ensuring that the original data sequence $x_{1:n}$ is always recoverable from $c(x_{1:n})$. 
Given a data generating source $\mu$, we know from Shannon's Source Coding Theorem that the optimal (in terms of expected code length) source code $c$ uses codewords of length $-\log_2 \mu(x_{1:n})$ bits for all $x_{1:n}$.
This motivates the notion of the \emph{redundancy} of a source code $c$ given a sequence $x_{1:n}$, which is defined as
$r_c(x_{1:n}) := \ell_c(x_{1:n}) + \log_2 \mu(x_{1:n})$.
Provided the data generating source is known, near optimal redundancy can essentially be achieved by using arithmetic encoding \citep{Witten87}.
More precisely, using $a_\mu$ to denote the source code obtained by arithmetic coding using probabilistic model $\mu$, the resultant code lengths are known to satisfy
\begin{equation}\label{eq:coding_redundancy}
\ell_{a_\mu}(x_{1:n}) < -\log_2 \mu(x_{1:n}) + 2,
\end{equation}
for all $x_{1:n}$, which implies that the redundancy is always less than $2$. 
In practice however, the true data generating source $\mu$ is typically unknown.
The data can still be coded using arithmetic encoding with an alternate coding distribution $\rho$, however now we expect to use an extra $\mathbb{E}_\mu \left[ \log_2 \mu(x_{1:n}) / \rho(x_{1:n}) \right]$ bits to code the random sequence $x_{1:n} \sim \mu$.
From here onwards, we restrict our attention to that of specifying a good coding distribution.

\section{Sparse Sequential Dirichlet Distribution}

We now propose an adapted version of the Sequential Dirichlet Distribution, which will use less computational resources than the Sequential Sub-Alphabet Estimator, while still performing well in situations where $|\cA|$ is much less than $|\cX|$.

\begin{defn}
Given an alphabet $\cX$, for all $n\in\mathbb{N}$ and for all $x_{1:n} \in \cX^n$, the Sparse Sequential Dirichlet distribution $\xi : \cX^* \to (0,1]$ is defined as
\begin{equation}
\xi(x_{1:n}) := \prod\limits_{i=1}^n \mathbb{I}[c(x_{1:i}) = 0] \, \alpha_i \frac{1}{|\cX| - |U(x_{<i})|} + 
\mathbb{I}[c(x_{1:i}) > 0] \, (1 - \alpha_i)  \frac{c(x_{1:i})+ \tfrac{1}{2}}{i+\tfrac{|U(x_{<i})|}{2}-1}
\end{equation}
where $c(x_{1:n}) := \sum_{i=1}^{n-1} \mathbb{I}[x_n = x_i]$, $U(x_{1:n}) := \{ s \in \cX : c(x_{1:n}s) > 0\}$ and $\alpha_i := \tfrac{1}{i}$ for $i \in \mathbb{N}$.
\end{defn}

In the above, $U(x_{1:n})$ is simply the number of distinct symbols occurring in $x_{1:n}$.
Furthermore, one can easily verify that $\xi$ is a valid probability measure over finite but arbitrarily large strings whose symbols are from the alphabet $\cX$.

\paragraph{Computational Properties.}

Given a sequence $x_{1:n} \in \cA^n$, $\xi(x_{1:n})$ can be computed in $O(n)$ time, with $O(|\cA|)$ space required to store the counts for the seen symbols.
Furthermore, by using $\xi(x_n \,|\, x_{<n}) = \xi(x_{1:n}) / \xi(x_{<n})$ in combination with the chain rule $\xi(x_{1:n}) = \xi(x_n \,|\, x_{<n}) \xi(x_{<n})$, each symbol $x_{n+1}$ can be processed in $O(1)$ time, leading to a straightforward incremental algorithm.
As usual, numerical underflow issues can be addressed by storing all probability values in log-space.

\paragraph{Analysis.}

We now show that Sparse Sequential Dirichlet Coding using an alphabet of $\cX$ performs well provided there exists an alphabet $\cA \subset \cX$ for which Sequential Dirichlet Coding performs well. 
Our goal will be to provide a redundancy bound which does not exhibit a linear dependence on $|\cX|$.

\begin{thm}
\label{thm:ssdc_main_result}
Given alphabets $\cX$ and $\cA$ such that $\cA \subseteq \cX$, for all $n \in \mathbb{N}$, for all $x_{1:n} \in \cA^n$, we have
$
-\log_2 \xi(x_{1:n}) \leq \log_2 n + |\cA| \log_2 |\cX| - \log_2 \rho_{\cA}(x_{1:n}).
$
\begin{proof}
First note that since $|\cX| \geq |\cX| - |U(x_{<i})|$ and $|U(x_{1:n})| \leq |\cA|$ for all $x_{1:n} \in \cA^n$,
\begin{equation*}
\xi(x_{1:n}) \geq \prod\limits_{i=1}^n \mathbb{I}[c(x_{1:i}) = 0] \, \alpha_i \frac{1}{|\cX|} + 
\mathbb{I}[c(x_{1:i}) > 0] \, (1 - \alpha_i)  \frac{c(x_{1:i})+ \tfrac{1}{2}}{i+\tfrac{|\cA|}{2}-1}.
\end{equation*}

Now, noting that $\alpha_i = \tfrac{1}{i} \geq \tfrac{1}{2} / (i + |\cA|/2 -1)$ for all $i \in \mathbb{N}$, we get
\begin{equation*}
\xi(x_{1:n}) \geq \prod\limits_{i=1}^n  \frac{c(x_{1:i})+ \tfrac{1}{2}}{i+\tfrac{|\cA|}{2}-1} \left( \mathbb{I}[c(x_{1:i}) = 0] \frac{1}{|\cX|} + 
\mathbb{I}[c(x_{1:i}) > 0] (1 - \alpha_i) \right).
\end{equation*}
Since there can be at most $|\cA|$ new symbols, with the first symbol always being new, and 
$$\prod_{\substack{1 \leq i \leq n \,:\, \mathbb{I}[c(x_{1:i}) > 0]}} \hspace{-2em} (1-\alpha_i ) \geq \prod_{i=2}^n (1-\alpha_i ),$$ 
we can conclude
\begin{equation}
\label{eq:bound_worker1}
\xi(x_{1:n}) \geq |\cX|^{-|\cA|} \prod\limits_{i=1}^n \frac{c(x_{1:i}) + \tfrac{1}{2}}{i + \tfrac{|\cA|}{2} - 1} \prod_{i=2}^n (1-\alpha_i ).
\end{equation}
Now, simplifying the telescoping product and applying the definition of $\rho_{\cA}$ (see Equation \ref{eq:sequential_dirichlet_def}) to the right-hand side of Equation \ref{eq:bound_worker1} gives
$
n^{-1} |\cX|^{-|\cA|} \rho_{\cA}(x_{1:n})
$.
Hence,
\begin{equation*}
-\log_2 \xi(x_{1:n}) \leq -\log_2 n^{-1} |\cX|^{-|\cA|} \rho_{\cA}(x_{1:n})
= \log_2 n + |\cA| \log_2 |\cX| - \log_2 \rho_{\cA}(x_{1:n}).
\end{equation*}
\end{proof}
\end{thm}
Thus, combining Theorem \ref{thm:ssdc_main_result}, Equation \ref{eq:coding_redundancy} and Equation \ref{eq:small_alphabet_dirichlet_bound}, the overall coding redundancy of the Sparse Sequential Dirichlet Distribution is upper bounded by
\begin{equation}\label{eq:ssdc_redundancy}
 \frac{|\cA| + 1}{2} \log_2 n + |\cA|\log_2 |\cX| + |\cA| + 1.
\end{equation}

\paragraph{Discussion.}

A comparison of Equation \ref{eq:ssdc_redundancy} to Equation \ref{eq:dirichlet_bound} suggests that the redundancy of Sparse Sequential Dirichlet Coding will be less than Sequential Dirichlet Coding when $|\cA|$ is much smaller than $|\cX|$. 
Furthermore, by applying the inequalities
$$|\cA| \log_2 \tfrac{|\cX|}{|\cA|} \leq \log_2 {|\cX| \choose |\cA|} \leq |\cA| \log_2 \tfrac{e |\cX|}{|\cA|}$$
to bound Equation \ref{eq:redun_seqsubalpha}, we can see that our redundancy bound for Sparse Sequential Dirichlet Coding is competitive with the redundancy bound for the Sequential Sub-alphabet estimator whenever $|X|$ is much larger than $|A|$, and worse when $|\cA|$ is close to $|\cX|$. 

\section{Numerical Experiments}

We now present some numerical results for Sparse Sequential Dirichlet Coding, by
comparing and contrasting our technique using the experimental framework described below.

\paragraph{Experimental Setup.}

Each different experiment consisted of evaluating the performance of 5 different coding distributions on synthetically generated data, for various choices of $\cA$ and $\cX$.
The first technique, {\sc oracle}, used the true underlying data generating distribution to code the data.
This is of course the optimal coding distribution in expectation, and a natural baseline.
The second and third techniques, {\sc sdc}$(\cA)$ and {\sc sdc}$(\cX)$, refer to using Sequential Dirichlet Coding using the alphabets $\cA$ and $\cX$ respectively.
These two methods allow us to measure the impact of knowing and not knowing $\cA$ in advance.
{\sc ssd} refers to our Sparse Sequential Dirichlet Coding technique.
Finally, {\sc ssa} refers to the Sequential Sub-Alphabet technique of \cite{Tjalkens93sequentialweighting}.

To evaluate each particular combination of $\cA$ and $\cX$, 100,000 parameter vectors, $\mathbf{a}_i\in\mathbb{R}^{|\cA|}$ for $1 \leq i \leq 100,000$, were sampled from a Symmetric Dirichlet Distribution using a concentration parameter of $1.0$.
These $\mathbf{a}_i$ were used to define a set of Categorical Distributions over the symbols in $\cA$.
Each Categorical Distribution was used once to generate a data sequence of $100$ independent and identically distributed random symbols, which were then coded using each of the methods.
The mean, min and max performance, measured in bits, for each different coding distribution on the generated data sequences was then summarised in Tables \ref{tbl:results1}, \ref{tbl:results3} and \ref{tbl:results2}.
Additionally, the last line of each table measured how many extra bits Sparse Sequential Dirichlet Coding needed compared with Sequential Sub-Alphabet Coding.

\paragraph{Results.}

\begin{table}[t!]
\centering
\begin{tabular}{ | c | c | c | c | }
\hline
Method  & Mean & Min & Max\\ 
\hline\hline
        {\sc oracle} &185.048&12.3267&244.107 \\
   {\sc sdc}$(\cA)$&193.953 &21.368&243.718 \\
       {\sc sdc}$(\cX)$ &236.343 &63.7581&286.108 \\
      {\sc ssd} &210.844 &23.7755&262.4 \\
       {\sc ssa} &212.257 &24.4074&262.928 \\
\hline\hline       
     {\sc ssd} - {\sc ssa} &-1.41272&-5.77022&0.366856 \\
\hline
\end{tabular}
\caption{\small{Number of bits needed to encode 100 symbols when $|\cA| = 5$ and $|\cX| = 26$.}}
\label{tbl:results1}
\end{table}
\begin{table}[t!]
\centering
\begin{tabular}{ | c | c | c | c | }
\hline
Method  & Mean & Min & Max\\ 
\hline\hline
           {\sc oracle}&278.363&131.529&340.359\\
   {\sc sdc}$(\cA)$&293.969&146.716&349.882\\
       {\sc sdc}$(\cX)$&492.284&345.031&548.197\\
      {\sc ssd}&349.169&181.365&412.766\\
       {\sc ssa}&350.473&187.604&411.656\\
\hline\hline       
     {\sc ssd} - {\sc ssa}&-1.30374&-7.4791&1.3234\\
\hline
\end{tabular}
\caption{\small{Number of bits needed to encode 100 symbols when $|\cA| = 10$ and $|\cX| = 256$.}}
\label{tbl:results3}
\end{table}

Table \ref{tbl:results1} and Table \ref{tbl:results3} compare the relative coding performance of Sparse Sequential Dirichlet Coding when $|\cA|$ is much less than $|\cX|$.
In both situations we see that the Sparse Sequential Dirichlet technique is on average slightly superior to the Sequential Sub-Alphabet method, and never worse by more than $1.32$ bits.
Both techniques performed significantly better than the Sequential Dirichlet Coding method which used the alphabet $\cX$.
This is consistent with the redundancy bounds we presented earlier.
Lastly, Table \ref{tbl:results2} gives an example of what can happen when the sparsity assumption doesn't apply.
Here we see that the Sparse Sequential Dirichlet method is outperformed by all other techniques, though not by a large margin.

\paragraph{Discussion.}
    
In light of its superior computational properties, our results suggest that the Sparse Sequential Dirichlet technique is a good alternative to the Sequential Sub-Alphabet method whenever $|\cA|$ is much less than $|\cX|$.
If issues of computation or limited memory are not an issue, the Sequential Sub-Alphabet method is to be preferred due to its better performance when $|\cA|$ is not much less than $|\cX|$.

\begin{table}[t!]
\centering
\begin{tabular}{ | c | c | c | c | }
\hline
Method  & Mean & Min & Max\\ 
\hline\hline
        {\sc oracle}&360.053&234.325&422.467\\
   {\sc sdc}$(\cA)$&382.911&258.392&440.005\\
       {\sc sdc}$(\cX)$&396.527&272.007&453.62\\
      {\sc ssd}&410.573&277.942&476.754\\
       {\sc ssa}&397.344&271.68&456.234\\
\hline\hline       
     {\sc ssd} - {\sc ssa}&13.2292&0.446927&20.5248\\     
\hline
\end{tabular}
\caption{\small{Number of bits needed to encode 100 symbols when $|\cA| = 18$ and $|\cX| = 26$.}}
\label{tbl:results2}
\end{table}

\section{Conclusion}

This short paper has described a simple and efficient coding technique for multi-alphabet memoryless sources.
It provably works well when only a small subset of possible alphabet symbols are expected to occur in any given data sequence.
As future work, it would be interesting to explore the applicability of this technique as a building block within more sophisticated context modeling techniques.

\paragraph{Acknowledgements}
The authors would like to thank Kee Siong Ng and Marc Bellemare for comments that helped improve this paper.
{
\bibliographystyle{plainnat}
\bibliography{ssdc}

\begin{thebibliography}{3}
\providecommand{\natexlab}[1]{#1}
\providecommand{\url}[1]{\texttt{#1}}
\expandafter\ifx\csname urlstyle\endcsname\relax
  \providecommand{\doi}[1]{doi: #1}\else
  \providecommand{\doi}{doi: \begingroup \urlstyle{rm}\Url}\fi

\bibitem[Krichevsky and Trofimov(1981)]{krichevsky1981pue}
R.~Krichevsky and V.~Trofimov.
\newblock {The performance of universal encoding}.
\newblock \emph{Information Theory, IEEE Transactions on}, 27\penalty0
  (2):\penalty0 199--207, 1981.

\bibitem[Tjalkens et~al.(1993)Tjalkens, Shtarkov, and
  Willems]{Tjalkens93sequentialweighting}
Tjalling~J. Tjalkens, Yuri~M. Shtarkov, and Frans M.~J. Willems.
\newblock Sequential weighting algorithms for multialphabet sources.
\newblock In \emph{6th Joint Swedish-Russian Int. Worksh. Inform. Theory},
  pages 22--27, 1993.

\bibitem[Witten et~al.(1987)Witten, Neal, and Cleary]{Witten87}
Ian~H. Witten, Radford~M. Neal, and John~G. Cleary.
\newblock Arithmetic coding for data compression.
\newblock \emph{Commun. ACM}, 30:\penalty0 520--540, June 1987.
\newblock ISSN 0001-0782.

\end{thebibliography}
}

\end{document}